\documentclass[11pt]{article}

\usepackage{amsfonts}
\usepackage{graphicx}
\usepackage{amsmath}

\makeatletter \@addtoreset{equation}{section}
\newcommand{\ii}{\'{\i}}

\newcommand{\be}{\begin{equation}}
\newcommand{\ee}{\end{equation}}
\newcommand{\bea}{\begin{eqnarray}}
\newcommand{\eea}{\end{eqnarray}}
\begin{document}

\title{
\begin{flushright}
 {\normalsize \small
GNPHE/0906 }
 \\[1cm]
 \mbox{}
\end{flushright}
\textbf{Extremal Black Brane Attractors }\\
\textbf{on The Elliptic Curve} }
\author{ \hspace*{-20pt} Rachid Ahl Laamara$^{1,4}$\thanks{%
doctorants.lphe@fsr.ac.ma}, Manuel Asorey$^{2}$\thanks{asorey@unizar.es},  Adil Belhaj$^{3,4}$\thanks{%
belhaj@unizar.es}, Antonio Segui$^{2}$\thanks{segui@unizar.es} \\
{\small $^{1}$Lab/UFR-Physique des Hautes Energies, Facult\'{e} des
Sciences, Rabat, FSR, Morocco}\\
{\small $^{2}$Departamento de F\ii sica Te\'orica, Universidad de
Zaragoza,
50009-Zaragoza, Spain}\\
{\small $^{3}$Centre National de l'Energie, des Sciences et des
Techniques
Nucleaires, CNESTEN }\\
[-6pt] {\small Cellule Sciences de la Mati\`ere, Rabat, Morocco }\\
{\small $^{4}$Groupement National de Physique des Hautes Energies, GNPHE, Si\`{e}ge focal: FSR}\\
[-6pt] {\small  Rabat, Morocco }} \maketitle

\begin{abstract}
Reconsidering  the analysis  of the moduli space of $N=2$ eight
dimensional supergravity coupled to seven scalars, we
propose a new scalar  manifold  factorization given by $\frac
{\textsc {SO(2,2)}}{{\textsc{ SO(2)}}\times {\textsc{ SO(2)}}}\times \frac{\textsc{ SO(2,1)}%
}{\textsc{ SO(2)}}\times \textsc {SO(1,1)}$. This factorization is  supported by  the
appearance  of three solutions
 of Type IIA extremal black $p$-branes $%
(p=0,1,2)$ with $AdS_{p+2}\times S^{6-p}$ near-horizon geometries in
eight dimensions. We analyze  the corresponding attractor mechanism.
In particular,  we give  an  interplay   between the scalar manifold
factors and the extremal black $p$-brane charges. Then we  show
that the dilaton can be stabilized by the dyonic black 2-brane
charges.

\end{abstract}


\thispagestyle{empty}

\newpage \setcounter{page}{1} \newpage

\section{Introduction}

In the last years, four dimensional extremal black hole
attractors have received an increasing attention in the context of
supergravity theories embedded in superstrings and M-theory compactified on Calabi-Yau
manifolds\cite{OSV}-\cite{CFM}.  In  the near-horizon geometry limit, some of the
supergravity  scalar fields, obtained from geometric deformations,
 can take fixed values in terms of
the black hole charges. Several studies of the $N=2$
attractor mechanism  in Type  IIA superstring on Calabi-Yau 3-folds
reveal that in  such a  limit  only the complexified K\"{a}hler moduli
 can be stabilized by the abelian black hole  charges.
 The  remaining scalar fields  corresponding  to the complex
structure deformations,
 the values of the \textbf{R-R} C-fields on 3-cycles, the
dilaton and the axion,  remain  free  and  can  take arbitrary values.

However, the situation in six dimensions, which is obtained from
Type IIA superstring on  the K3 surface, is somewhat different. In
this model,  the  attractor mechanism deal with the full  geometric
moduli space including    both the complexified K\"{a}hler and the
complex structure deformations \cite{BDSS,S1,SS,B}. These  geometric
parameters can be combined with  the \textbf{NS-NS} B-field values,
on non trivial 2-cycles of the K3 surface,  to form   a quaternionic
scalar manifold. Using a matrix formulation for   such a
quaternionic part, the geometric moduli of the K3 surface and the
value of the \textbf{NS-NS} B-field are determined by the abelian
black hole charges. However, the dilaton, being identified with a
non compact orthogonal direction (SO(1,1)), is attracted at the
horizon of the  extremal black F-string in terms of its charges
(electric and magnetic).

More recently, a special effort has been devoted to  discuss
extremal black branes in higher dimensional  supergravities
\cite{FMMS}. This concerns the intersecting attractors involving
extremal black holes and black strings in $D>5$ dimensional
supergravities.  In particular,    effective potentials and entropy
functions have been computed  in terms of U-duality black brane
charge invariants.

The main purpose of this paper is  to reanalyze  the moduli space of
eight dimensional  supergravity with $N=2$ supersymmetry.  We study
the attractor mechanism in the framework of  Type IIA superstring on
the elliptic  curve.  In particular we consider the extremal black
$p$-branes. The corresponding  near-horizon geometries are  given by
products of $AdS_{p+2}$ with ${(6-p)}$-dimensional real spheres
$S^{6-p}$. In  this work, we propose a new realization for the
scalar manifold of $N=2$ supergravity in eight dimensions. This
moduli space realization involves three factors  which correspond to
the appearance  of three solutions of eight dimensional extremal
black $p$-branes with $p=0,1,2$. Motivated  by the compactification
on the K3 surface, we discuss the attractors of such extremal black
objects. In particular,  we  establish a correspondence between the
scalar manifold factors and the extremal black objects  charges.
Then, a special interest is devoted to the dyonic black 2-brane
attractor. The minimum of its effective potential gives the value of
the dilaton in the $AdS_4\times S^4$ near-horizon geometry limit.

The organization of this paper is as follows. In section 2,  we develop
a new factorization  for  the scalar manifold of $N=2$ eight
dimensional supergravity involving three factors. The
identification of each factor is based  on the appearance of three
different extremal black $p$-branes. In section 3, we
discuss the attractor mechanism of such  black objects.
In particular, we point out  the existence of a link between the extremal black
brane charges and the $N=2$ scalar manifold factors in eight dimensions.
The last section is devoted to discussion of the results and  open questions.

\section{On The Moduli Space of $N=2$ Supergravity in Eight Dimensions}

In this section we reconsider  the analysis of  the scalar manifold of
$N=2$  supergravity in eight dimensions.  The  embedding of extremal black
$p$-branes  in Type IIA superstring  theory on the elliptic curve
 provides  a new factorization scheme for the seven scalars.
 We begin first by a  review on the  study of the scalar manifold.
We  will restrict ourself   to the bosonic sector. The  field content of the model
consists of a graviton, six vector fields, three 2-form gauge
fields, one self dual 3-form  gauge field and seven scalars. This
spectrum can be obtained from the  reduction  of M-theory by $T^{3}$
with the $ \textsc{ SL(3,R)}\times \textsc{ SL(2,R)}$ U-duality group \cite{SSb}. The
 full  spectrum is  given by
\begin{equation}
(G_{\mu \nu },B_{\mu \nu }^{I},A_{\nu }^{I\alpha },C_{\mu \nu \rho
},L_{I}^{\Lambda },L_{i}^{\alpha }),\quad \Lambda ,I=1,2,3\quad
\alpha, i =1,2.
\end{equation}
$L_{\alpha }^{I }$ is the coset representative of  $\frac{\textsc{ SL(3,R)}}{
\textsc{ SO(3)}}$,  while $ L_{i}^{\alpha }$ is the coset representative of  $\frac{ \textsc {SL(2,R)}%
}{ \textsc{ SO(2)}}$. In the M-theory picture,   the scalar fields are the coordinates of the following homogeneous space
\begin{equation}
\frac{\textsc {SL(3,R)}}{\textsc {SO(3)}}\times \frac{\textsc{ SL(2,R)}}{\textsc {SO(2)}}.
\end{equation}
This scalar manifold  represents  the deformation of the metric and
the value of the eleven dimensional supergravity 3-form  on
$T^{3}$.  Up  to an overall scale factor  corresponding to the size
of $T^{3}$, $\frac{\textsc{ SL(3,R)}}{\textsc {SO(3)}} $ determines the choice of the
metric of $T^{3}$. $\frac{\textsc{ SL(2,R)}}{\textsc {SO(2)}}$ is  coordinated  by  the volume
of $T^{3}$ and the value the 3-form gauge field takes on it.

Because  of the  strong coupling limit   duality, the spectrum of the
above eight dimensional supergravity can  also be obtained from the
reduction of  Type IIA superstring by $T^{2}$ (that is an $ S^1\times S^1$ fibration).
This manifold   has an obvious flat metric  which  depends on the size  of the   two circles.
The moduli  space of $T^{2}$ contains   a real parameter  describing
 its size   and a  complex parameter controlling its shape. The compactification of  the
massless bosonic   ten dimensional  Type IIA superstring  fields
\begin{equation}
\mathbf{NS-NS}:G_{MN},\;\;B_{MN},\;\;\phi \qquad \mathbf{R-R}
:A_{M},\;\;C_{MNK}  \quad  M,N,K=0,\dots,
\end{equation}
 gives the same spectrum discussed  above.

Motivated by the  results obtained from the K3-attractors, and based on the T-duality  groups in
Type II superstrings, we  propose here a new
factorization for the scalar manifold of $N=2$ supergravity in eight dimensions  involving three factors.
 This factorization will be related  to the   appearance of three extremal black  $p$-brane solutions
 with  $p=0,1,2$. To do so, let us first  recall  some ideas about the K3-attractors. The
corresponding  $N=2$ supergravity  consists of, among other, $(4\times 20)+1=81$ scalar fields, a
$ \textsc{U(1)}^{24}$ gauge symmetry and one self dual antisymetric  B-field.
The scalars span  the  following  manifold product
\begin{equation}
\label{mk3}
\frac{\textsc {SO(4,20)}}{\textsc{SO(4)}\times\textsc{SO(20)}}\times \textsc{SO(1,1)}.
\end{equation}
Here  $\frac{ \textsc{ S0(4,20)}}{ \textsc{ SO(4)}\times  \textsc{ SO(20)}}$ describes  the geometric
moduli
 space admitting a hyper-Kahler structure.  The extra factor $ \textsc{ SO(1,1)}$  stands for the
  dilaton, defining the string coupling constant $g_s$. It turns out that the two coset factors
 appearing in (\ref{mk3}) corresponds  to two different extremal black objects in six dimensions required by the electric/magnetic duality.
 If we have an electrically
charged $p$-brane, the magnetically charged dual object is a $q$-brane such that
\begin{equation}
p+q=2.  \label{d6d}
\end{equation}%
There are essentially two  six dimensional  extremal black $p$-brane
solutions defined by $p=0,1$ with $AdS_{p+2}\times S^{4-p}$
near-horizon geometries. $p=0$ corresponds to $AdS_{2}\times S^{4}$
describing an electric charged black hole, dual to the magnetic
black 2-brane with $AdS_{4}\times S^{2}$  near-horizon geometry.
$p=1$ is associated with a dyonic
black F-string  whose  near horizon limit is  $AdS_{3}\times S^{3}$. From the study of the attractor horizon geometries
of extremal  black $p$-branes ($p=0,1$) in Type IIA superstring on
the $K3$ surface \cite{BDSS}, we obtain  the following connection
\begin{table}[tbph]
\centering
\begin{tabular}{|c|c|c|}
\hline Coset space & Black brane & Gauge symmetry  \\ \hline
$\frac{ \textsc{ S0(4,20})}{ \textsc{SO(4)}\times \textsc{SO(20)}}$ & black holes (black 2-branes)
& $ \textsc{  \textsc{U(1)}}^{24}$ \\ \hline $ \textsc{SO(1,1)}$ & black F-string & $ \textsc{  \textsc{U(1)}}$ \\
\hline
\end{tabular}%
\caption{This table gives  the correspondence between the scalar
manifold factors and extremal black $p$-brane charges in Type IIA
superstring on K3 surface.} \label{table1}
\end{table}

Motivated by these  results, we expect to have a similar
correspondence in the $N=2$ eight dimensional supergravity theory embedded in  Type IIA
superstring  compactified on $T^{2}$. In this way,  the analogue of the equation (\ref{d6d}) reads
as
\begin{equation}
p+q=4,   \label{d8d}
\end{equation}%
and can be solved in three different   ways like

\begin{table}[tbph]
\centering
\begin{tabular}{|c|c|}
\hline $p=0\;\;(q=4)$  &  \text{Black holes}\;(\text{dual black
4-branes})  \\ \hline $p=1\;\;(q=3)$ &  \text{Black
strings}\;(\text{dual
black 3-branes}) \\ \hline $p=2\;\;(q=2)$ & \text{Dyonic black 2-branes} \\
\hline
\end{tabular}%
\end{table}
\vspace{2.7cm}

The extremal near-horizon geometries of these objects are  given by
the products of $AdS_{p+2}$ with real spheres $S^{6-p}$. In eight dimensions, they  are classified  into three categories.
\newline
\textbf{\small 1}. $p=0$ corresponds to an $AdS_{2}\times S^{6}$
 describing the  near-horizon geometry  of  electric charged  black holes.
 Their  dual magnetic are  black   4-branes with $AdS_{6}\times S^{2}$ near-horizon geometries.
The  objects  carry charges associated with the gauge invariant field
strengths $F^i=dA^i(i=1,\ldots,6)$ of the $N=2$ supergravity theory.\newline
\textbf{\small 2}. $p=1$ is associated with  $AdS_{3}\times S^{5}$
describing near-horizon geometry  of  extremal
 black strings.  They carry  electric charges  associated
with  3-form field strengths $H^i=dB^i(i=1,\ldots,3)$. The electric charge is
proportional to the integral of $\star H$ over $S^{5}$   that encloses
the string. The magnetic dual horizon geometry reads as
$AdS_{5}\times S^{3}$ and describes black 3-branes. They are charged
under the gauge invariant 3-form field strengths $H^i=dB$.
\newline \textbf{\small 3}. $p=2$  corresponds to a dyonic black 2-brane  with $AdS_{4}\times S^{4}$
 near-horizon geometry. This object
carries  both electric  and  magnetic charges associated with the single
gauge invariant 4-form field strength $G=dC$. \newline  We shall see that
this classification could be related to the moduli space of Type IIA
superstring on Calabi-Yau spaces. In this compactification, there are  three different
contributions  classified as follows:

\begin{itemize}
\item The dilaton defining the string coupling constant.

\item The geometric deformations of the Calabi-Yau space including the
antisymmetric B-field of the \textbf{NS-NS} sector. Depending on the Calabi-Yau spaces, these parameters  involve
complex structure deformations,  complexified K\"{a}hler deformations or
both.
\item The scalar moduli described by  the values of the  \textbf{R-R} gauge
 fields wrapped on non trivial
cycles in the  Calabi-Yau spaces.
\end{itemize}
According to this   observation and  based  on the K3-attractor results
mentioned earlier, the moduli space of Type IIA superstring  compactified on $T^{2}$ should have a priori three factors.
  They  may be related
 to the  existence of three  solutions  of extremal black $p$-branes with $p=0,1,2$.
 Under this hypothesis, the corresponding scalar manifold  should
take the form
\begin{equation}
M_{1}\times M_{2}\times M_{3}.  \label{fac}
\end{equation}
The identification of each factor can be obtained
by the help of the conjecture given in \cite{B}. This conjecture  can  be  re-formulated  in the framework of the
elliptic curve  as
follows:\newline \textbf{\small 1}. The black hole  charges fix only
the geometric deformations of the elliptic curve (the complex structure deformation, the  K\"{a}hler
deformation and the \textbf{NS-NS} B-field    on $T^2$).
\begin{equation*}
\begin{array}{ccc}
p=0 & \quad \text{Black hole charges}\rightarrow & \text{The geometric parameters.%
}%
\end{array}%
\end{equation*}
\textbf{\small 2}. The moduli related with  the values of the
\textbf{R-R} gauge vectors  on  1-cycles of the elliptic curve should be
fixed by black  string charges
\begin{equation*}
\begin{array}{ccc}
p=1 & \quad \text{Black string charges}\rightarrow & \text{ The \textbf{%
R-R} stringy moduli.}%
\end{array}%
\end{equation*}
\textbf{\small 3}. The dilaton could be fixed as usuall by the
dyonic state, which in this case is a black 2-brane
\begin{equation*}
\begin{array}{ccc}
p=2 & \quad \text{ Dyonic black 2-brane charges }\rightarrow &
\text{ the
dilaton.}%
\end{array}%
\end{equation*}
Based on this  conjecture, we will  identify  each $M_i$  factor of
(\ref{fac}) with  a coset space.
The T-duality  role in  Type IIA superstring on the elliptic curve
requires that one factor should be identified with
\begin{equation*}
M_{2}=\frac{ \textsc{ SO(2,2)}}{\textsc{SO(2)}\times\textsc{SO(2)}},
\end{equation*}%
determined by $2\times 2=4$ parameters  controlling  the metric  deformation   and the value of the \textbf{NS-NS} B-field on of the
elliptic curve. The T-duality
group of Type II superstrings  is $ SO(2,2)$ and that of M-theory is $SL(3)$. This result can be generalized to the compactification on $T^d$. In this case, the geometric scalar fields parametrize the coset space $\frac{ \textsc{ SO(2,2)}}{\textsc{SO(2)}\times\textsc{SO(2)}}$. This space corresponds to the choices of the metric with $\frac{d(d+1)}{2 }$ degrees of freedom and the values of the antisymmetric B-field with $\frac{d(d-1)}{2 }$ contributions. \\
The dilaton is related to the string coupling
\begin{equation*}
g_s\sim exp(\phi).
\end{equation*}
This is invariant under the shift $\phi \to \phi+\alpha$. At some points of the moduli space of $T^2$, $\alpha$ can be related to
the complexified K\"{a}hler parameter. In fact,   it is related to the Riemanian  volume of $T^2$ and the volune provided by the B-field. These two parameters are related by
the $SO(1,1)$  group. In this way, the dilaton in eight dimensions  can be represented, as in six dimensions, by a non compact
circle given by
\begin{equation*}
M_{1}= \textsc{ SO(1,1)}.
\end{equation*}
The two remaining parameters specifying the Wilson line  on $T^{2}$ can
 be combined into a
complex field. It  can  be  represented  by  the  coset space
\begin{equation*}
M_{3}= \frac{ \textsc{ SO(2,1)}}{ \textsc{ SO(2)}}.
\end{equation*}%
Finally, the factorization (\ref{fac})   reads  as
\begin{equation}
\frac{ \textsc{SO(2,2)}}{ \textsc{SO(2)}\times  \textsc{SO(2)}}\times \frac{%
 \textsc{SO(2,1)}}{ \textsc{SO(2)}}\times  \textsc{SO(1,1)}.
\end{equation}
It is worth while to comment a possible interrelation between  this Type superstring
factorization and the M-theory one. The factor $\frac{ \textsc{ SO(2,2)}}{\textsc{SO(2)}\times\textsc{SO(2)}}$
can be  parametrized by two complex scalar fields. One of them can be identified with the complexified volume
 of $T^3$ of the M-theory compactification given by the $\frac{ \textsc{SL(2,R)}}{\textsc{SO(2)}}$. The other
  complex scalar  come from the  $\frac{\textsc{ SL(3,R)}}{\textsc {SO(3)}} $  coset space.  The  $\textsc{SO(1,1)}$
   factor   can also be deduced  from the last coset space. The  remaining two scalar fields of
    $\frac{\textsc{ SL(3,R)}}{\textsc {SO(3)}} $ can  be combined  in the    factor
     $\frac{ \textsc{ SO(2,1)}}{ \textsc{ SO(2)}}$.\\
Next we will analyze the  extremal black $p$-brane attractors which correspond to
this new realization of the scalar manifold of $N=2$
supergravity in eight dimensions.

\section{Attractor Mechanism on The Elliptic Curve}
Here we discuss  the  attractor mechanism  of extremal black
branes that appear in Type IIA superstring compactified  on the elliptic
curve. We start  by   briefly recalling  the main results   obtained in the context of
higher dimensional Calabi-Yau backgrounds. Consider Type IIA superstring on a Calabi-Yau
$n$-folds with $SU(n)$ $(n>0)$ holonomy group.In the low energy limit,  it leads to  supergravity
 models with only  $2^{6-n}$ supercharges.  The  near-horizon geometries
  of the extremal black $p$-branes dealt with here  are given by the product
 of $AdS$ spaces and  real spheres as follows
\begin{equation}
AdS_{p+2}\times S^{8-2n-p}.
\end{equation}%
The numbers $n$ and $p$ satisfy the constraint
\begin{equation}
2\leq 8-2n-p.  \label{c}
\end{equation}%
For the compactification on  the Calabi-Yau $n$-folds, the electric/magnetic duality
relating electric black $p$-branes to $q$-dimensional magnetic
ones reads as
\begin{equation}
p+q=6-2n.  \label{gd}
\end{equation}
In the case of the Calabi-Yau 3-folds,  this equation  can be solved   by $p=q=0$ describing  a dyonic black hole
with $AdS_{2}\times S^{2}$ near-horizon geometry. The corresponding compactification
 leads to a four dimensional
 $N=2$ supergravity with eight supercharges,  coupled
to a $ \textsc{  \textsc{U(1)}}^{h_{1,1}+1}$ abelian  symmetry \footnote{ $h_{1,1}$ is
the dimension of the complexified K\"{a}hler moduli space in Type
IIA superstring on Calabi-Yau 3-folds.}. There are also  scalars belonging
to the vector multiplets and hypermultiplets. The complexified K\"{a}hler
moduli space is associated with  the vevs of the vector multiplet
scalars.  The complex structure deformations, the \textbf{R-R}
gauge field contributions, the axion and the dilaton belong to the
hypermultiplets.  The scalars of the hypermultiplets
 do not play any role in the study of four dimensional
 black hole attractors and can be  ignored. In Type IIA superstring on Calabi-Yau 3-folds, several
studies    concerning  the  attractor  mechanism,  show    that in the $AdS_{2}\times
S^{2}$ near-horizon  limit  only the complexified K\"{a}hler moduli
can be fixed by the abelian black hole charges \cite{FK1}-\cite{Da}.
This can be obtained by minimizing the black hole  effective potential \cite{TT}.
 This potential appears in the action
of the $N=2$ supergravity  coupled to the Maxwell theory in
which the abelian gauge vectors come from the reduction of the
 3-form on 2-cycles of the Calabi-Yau 3-folds. The hypermultiplet scalars remain free and   take arbitrary values
  near the horizon limit of black holes. It is worth   to point out  that   the complex
structure parameters have been fixed in the context
of Type IIB superstring on the  Calabi-Yau 3-folds.

However, the situation in six dimensions is somewhat different. It
is obtained by  the  compactification  on the K3 surface   which is a Calabi-Yau 2-folds.
 This manifold  has a mixed geometric moduli
 space involving both the  K\"{a}hler and the complex structure deformations. These
scalars together with the values of the  \textbf{NS-NS} B-field are
fixed by the abelian charges of the extremal  black holes  with
$AdS_{2}\times S^{4}$ near-horizon geometry. In the $AdS_{3}\times S^{3}$ near-horizon  limit, the dilaton
has been fixed by the charges (electric and magnetic) of thr extremal
black F-string\cite{BDSS}.

So far we have recalled    lower dimensional  results;  now we consider   the case of $n=1$
corresponding to the reduction on  the elliptic curve.  The analysis
we follow here is quite similar to the case of $N=(1,1)$ supergravity in six dimensions
 obtained from the compactification on
the  K3 surface.  A close inspection reveals that the  six and eight
dimensions share  some similarities. In both  cases, they   involve
extremal black extended objects. This objects   are
 absent in the higher dimensional
Calabi-Yau compactification $n\geq 3$ \footnote{This can be easily  seen
from the equation (\ref{c}).}. Moreover,  the metric deformations and the
values of the  B-fields can be
collected in one coset space. They are given  by $\frac{\textsc {SO(4,20)}}{%
\textsc {SO(4)}\times\textsc{SO(20)}}$ for the K3
surface and $\frac{\textsc {SO(2,2)}}{\textsc{SO(2)}\times\textsc{SO(2)}}$ for $T^2$.

\subsection{Scalar Manifold Factors/ Extremal Black $p$-brane Charges Correspondence}
Here we analyze  the correspondence between the above  moduli
space factors and the $p$-brane charges.
In  eight dimensional $N=2$ supergravity,
the total abelian gauge  group    is
\begin{equation}
 \textsc{U(1)}_{b2b}\times  \textsc{U(1)}_{bs}^{3}\times  \textsc{U(1)}_{bh}^{6}.
\end{equation}%
The $ \textsc{U(1)}_{b2b}$ factor is the abelian gauge symmetry associated
with the single field strength of the 3-from, the \textbf{R-R} 3-field
which can be decomposed into a self dual  and anti-self dual
sectors leading to  the electric and magnetic charges of the dyonic
black 2-branes. The electric and magnetic charges can be rotated by  $ \textsc{ SO(1,1)}$ isotropy symmetry, and
 therefore,  it  corresponds to  the dilaton scalar manifold. \newline The
abelian factor $U(1)^{3}_{bs}$ corresponds to three  gauge   field
strengths $H^{i}=dB^{i}$ ($i=1,2,3$). One of them comes from the  \textbf{NS-NS}
sector associated  with the F-string,  while the two other
are obtained from the reduction of the \textbf{R-R} C-field on the two  1-cycles of
the elliptic curve. In this way, $ \textsc{U(1)}_{bs}^{3}$ gauge symmetry can be factorized as
\begin{equation}
 \textsc{U(1)}_{bs}^{3}= \textsc{U(1)}\times  \textsc{U(1)}^{2}.
\end{equation}%
This separation of the charges is governed by a $ \textsc{SO(2,1)}$ isotropy
symmetry.  A simple  inspection suggests  that this sector can  be
associated with the  coset space  $\frac{ \textsc{ SO(2,1)}}{ \textsc{SO(2)}}$.\newline
The last  factor $ \textsc{U(1)}_{bh}^{6}$ is the abelian gauge symmetry
associated with the  six field strength 2-forms $F^{i}$($i=1,\ldots,6$) of the eight
dimensional supergravity multiplet. These vector fields  arise not only from the
\textbf{R-R} sector, as in the case of higher dimensional Calabi-Yau
compactification, but also from the \textbf{NS-NS} sector due to the
fact that $b_{1}(T^{2})\neq 0$. One might  factorize  the $ \textsc{U(1)}_{bh}^{6}$   gauge
symmetry as
\begin{equation}
 \textsc{U(1)}_{bh}^{6}= \textsc{U(1)}^{2}\times  \textsc{U(1)}^{2}\times  \textsc{U(1)}\times  \textsc{U(1)}.
\end{equation}
These  abelian  gauge fields  can be obtained from the  \textbf{NS-NS} and  \textbf{R-R} sectors.
 The  $ \textsc{U(1)}^{2}\times  \textsc{U(1)}^{2}$ gauge sub-group are   obtained from the \textbf{NS-NS} sector.
  This part has only  $ \textsc{ SO(2)}\times  \textsc{ SO(2)}$ isotropy symmetry. There are another
  $ \textsc{U(1)}\times  \textsc{U(1)}$ gauge factor   which come    from  the
\textbf{R-R} sector. It is worth to recall that  these gauge fields are the {\bf  R-R} 1-form
 and the reduction of the 3-form on the
elliptic curve. With the presence of these fields,   we shall  see
that there is an enhancement of the isometry group. This can be
supported by the M-theory  uplifting  scenario. From M-theory  point
of view, the six gauge fields can be
 classified into  two categories. Three of them are obtained from the metric, while the remaining
 three one come from the the 3-form.  In fact,  the eleven dimensional metric
 gives two   vectors  belonging  to  the  \textbf{NS-NS} sector and one vector  living in the  \textbf{R-R}
 sector.  The corresponding  abelian  gauge charges can be
rotated by the $ \textsc{ SO(2,1)}$  isotropy symmetry.  The bosonic
field content of the reduction of the eleven dimensional  3-form
consists  of   a similar vector contributions  which   are also
rotated by $ \textsc{ SO(2,1)}$. The full isometry group should be $
\textsc{ SO(2,1)} \times  \textsc{ SO(2,1)}$. It is a well-known
fact that $ \textsc{ SO(2,1)} \times  \textsc{ SO(2,1)}$ is
equivalently  to $ \textsc{ SO(2,2)}$, as we want.  In this way, the
above   six charges can  be related to the  variation of the metric
and the B-field  on the elliptic curve ($\frac{ \textsc{ SO(2,2)}}{
\textsc{ SO(2)}\times  \textsc{ SO(2)}}$). This can be understood
from the equivalence
\begin{equation*}
\frac{ \textsc{ SO(2,2)}}{\textsc{SO(2)}\times\textsc{SO(2)}}\sim
\frac{ \textsc{ SO(2,1)}}{ \textsc{ SO(2)}}\times \frac{ \textsc{
SO(2,1)}}{ \textsc{ SO(2)}}.
\end{equation*}
Finally, we propose the following correspondence
\begin{table}[tbph]
\centering
\begin{tabular}{|c|c|c|c|}
\hline Coset space & Black objects & Gauge symmetry  \\ \hline
$\frac{ \textsc{ SO(2,2)}}{ \textsc{ SO(2)} \times \textsc{ SO(2)}}
$ & black holes (black 4-branes)
& $ \textsc{U(1)}_{bh}^{6}$ \\ \hline $\frac{ \textsc{ SO(2,1)}}{ \textsc{ SO(2)}}$ & black strings (black 3-branes) & $%
 \textsc{U(1)}_{bs}^{3}$ \\
\hline $ \textsc{ SO(1,1)}$ & black 2-brane & $ \textsc{U(1)}_{b2b}$\\
\hline
\end{tabular}%
\caption{This table describes the relation between scalar
manifold factors and extremal black $p$-brane charges in eight
dimensions.} \label{table2}
\end{table}
\vspace{2.7cm}
\subsection{Dyonic 2-brane Attractors}
Since the technical analysis of six and eight dimensions are quite
similar, we  shall only discuss  here   the dyonic
solution\footnote{ We hope to give a general solution in a  future
work \cite{ABDDJSS}.}. The  dyonic black 2-brane is associated with
the case of  $n=1$ and $p=2$ whose  near-horizon geometry is given
by  $AdS_{4}\times S^{4}$. In this configuration, there  are no
closed 2 and 3-forms, so the corresponding charges  are not allowed.
There are only 4-form charges (electric and magnetic) supported by  this geometry.
 Using an   analysis  similar to the one given in  \cite{FMMS}, the near-horizon
geometry ansatz of this configuration can  be written as
\begin{equation}
ds^2=r^2_{AdS}ds^2_{AdS_4}+r_S^2ds^2_{S^4}, \qquad
G_4=p\;\alpha_{S^4}+e\;\beta_{Ads_4},
\end{equation}
where  $\alpha_{S^4}$   and $\beta_{AdS_4}$ denote respectively the
volume forms of $S^4$  and  $AdS_4$. The magnetic charge $p$  and
the electric charge $e$  are defined by
\begin{equation}
p=\int_{S^4}G_4\qquad  e=\int_{AdS_4}\star G_4.
\end{equation}
 It is useful to introduce  the parameterization $
Q^{1,2}=\frac{1}{2}(p\pm e)$, so the central charges take the  following general form
\begin{equation}
Z=MQ,
\end{equation}
where $M$ is a  $2\times 2$ matrix parameterizing the $SO(1,1)$
factor. This matrix  is represented by
\begin{equation}
\label{matrix}
\left(%
\begin{array}{cc}
\cosh(2\phi) & \sinh(2\phi) \\
\sinh(2\phi) & \cosh(2\phi)
\end{array}%
\right),
\end{equation}
where $\phi$ is the dilaton scalar field. The dyonic black 2-brane  effective potential for the dilaton
 is given as usual in terms of the  central
charges. Using the equation  (\ref{matrix}), this  potential
reads as
\begin{equation}
V_{eff}=\frac{1}{2}(p^2\exp(-4\phi)+e^2\exp(4\phi)).
\end{equation}
The dilaton is stabilized  at the minimum of the previous potential
by  the following attractor equations
\begin{equation}
\frac{dV}{d\phi}=0, \qquad \frac{d^2V}{d^2\phi}>0.
\end{equation}
The solution of these equations is
\begin{equation}
\exp(4\phi)\sim\frac{p}{e}.
\end{equation}
We obtain here   exactly the same value of the dilaton
that appears in the K3 attractor.
 In both cases the dilaton can be fixed by the electric and magnetic charges of the dyonic object.  The only difference
 that occur  in eight dimensions  is that the dilaton can  be
  fixed by black 2-brane, while in six dimensions it has been fixed by black string  charges.

\section{Conclusions and open questions}

We have  reconsidered the analysis of the moduli space  of $N=2$
eight dimensional  supergravity. The near-horizon geometries of the
corresponding extremal black $p$-branes have been assumed to be
products of $AdS_{p+2}$ with ${(6-p)}$-dimensional real spheres
$S^{6-p}$. Inspired from    K3-attractors, we have  proposed
 a new three factor realization for the scalar
manifold of  such a  $N=2$ supergravity model. This form  is based
on the existence   of three different  extremal black $p$-brane
solutions with $p=0,1,2$.  Using the conjecture introduced in
\cite{B},
 we have identified the black objects associated to each factor. In
particular, each coset space has been associated with a eight
dimensional black brane charge solution. The novel feature  is that
in this  case the number of U(1)  charges associated to the black
objects is larger than the dimension of the scalar factor moduli.
However, given the discrete character of the charges this effect,
which is not present in the six-dimensional case, does not imply by
any means any redundancy or over counting.

We have also  analyzed the attractor mechanism on the elliptic
curve,  and pointed out a correspondence between the scalar manifold
factors  and  the extremal black object charges. We specially
focused on the case of dyonic attractors. In the $AdS_4\times S^4$
near-horizon limit of the black 2-brane we have shown that upon
minimization the effective potential fixes  the value of the dilaton
in terms of electric and magnetic charges.

An interesting open question concerns   the solution of the
attractor equations for general extremal   black $p$-branes in eight
dimensions. It should be also interesting to look
 for non supersymmetric attractor solutions on the elliptic curve.

Another open problem is the analysis of extremal  black brane
attractors on general Riemann surfaces. In particular the study
would be interesting  for a two dimensional sphere $S^2$, where
there is  only one K\"{a}hler parameter controlling its size, and
the  gauge vectors come only  from the  \textbf{R-R} sector. In this
case, we guess that the size of  $S^2$ might be fixed by the
\textbf{R-R} black hole charges. We shall address these  open
questions  in the future \cite{ABDDJSS}.

\newpage
\textbf{Acknowledgments.}

We thank    B. Belhorma,
  L. J. Boya,  P. Diaz,  L.B. Drissi, H. Jehjouh,  S. Montanez,  J. Rasmussen, E. H. Saidi for
 collaborations, discussions on related subjects, correspondence and
scientific helps. This work has been   supported  by  the PCI-AECI (grant  A/9335/07), CICYT (grant
FPA-2006-02315) and DGIID-DGA (grant 2007-E24/2).

\end{document}